\documentclass{aa}  
\usepackage{graphicx}
\usepackage{booktabs}
\usepackage{tabularx} 
\usepackage{makecell} 
\usepackage{xcolor}%
\usepackage{graphicx,subcaption,lipsum}
\usepackage{threeparttable}
\usepackage{txfonts}
\usepackage[colorlinks=true,citecolor=blue,linkcolor=blue,urlcolor=blue]{hyperref}

\begin{document} 

\title{Shaken, not stirred: inefficient mixing of CM- and CI-like materials}

   \author{S.E. Anderson,
          \inst{1,2}
          P. Vernazza,
          \inst{2}
          \and
          M. Brož\inst{3}
          }

   \institute{CNRS, Observatoire de la Côte d’Azur, Laboratoire J.-L. Lagrange, CS 34229, 06304 Nice Cedex 4, France \\
              \email{sarah.anderson@oca.eu}
         \and
         LAM, Laboratoire d’Astrophysique de Marseille, Aix Marseille University, 38 rue Frédéric Joliot-Curie, 13013, Marseille, France  \\
              \email{pierre.vernazza@lam.fr}
         \and
             Charles University, Faculty of Mathematics and Physics, Institute of Astronomy, V Holešovičkách 2, CZ-18200, Czech Republic\\
             \email{mira@sirrah.troja.mff.cuni.cz}
             }

   \date{Received XXX; accepted YYY}

 
  \abstract
   {
A recent study suggests that CM chondrite-like planetesimals formed in the vicinity of Saturn, in a pressure bump outside the gap carved by proto-Jupiter. While a fraction of these objects was implanted into the asteroid belt as a consequence of Saturn's growth, it remains unclear whether the scattered remainder could reach the ice-giant region and mix with more distant carbonaceous reservoirs.
   }
   {
We test whether outward scattering during Saturn’s growth and migration can implant CM-like bodies onto long-lived orbits in the Uranus-Neptune region, where they could contaminate the CI reservoir.
   }
   {
We performed N-body integrations of 100-km planetesimals launched from the outer edge of Jupiter’s gap, including gas drag and the gravitational perturbations of growing Jupiter and Saturn, with optional inclusion of a nearby ice-giant embryo. We explored a range of gas surface-density profiles and growth timescales.
   }
   {
While Saturn’s growth efficiently scatters CM-like planetesimals, fewer than $\sim2\%$ are implanted beyond 15 au, even under gas-rich conditions, because gas drag damps their eccentricities and drives them back toward their pericentres rather than allowing them to circularize at larger distances. Adding an ice-giant core modestly increases the outward reach (up to ${\sim4}\%$ in the most gas-rich case), but Type-I migration further lowers perihelia, making long-term retention at large distances difficult. For a CM mass budget $M_{\rm CM,tot}\sim1\,M_\oplus$, this implies at most $M_{\rm CM}\lesssim0.02$–$0.04\,M_\oplus$ reaches 15-25\,au, corresponding to a diluted mass fraction $\lesssim(1$–$2)\times10^{-3}$ in the outer ring, hence negligible contamination of the CI reservoir.
   }
   {
Combined with the distinct radial distributions of CM- and CI-like asteroids in the belt, these results imply limited mixing of carbonaceous reservoirs and isolation of the CI reservoir. This strongly suggests that Uranus and Neptune formed later than Jupiter and Saturn, under lower gas densities, supporting a sequential giant-planet formation model.
}

   \maketitle
\nolinenumbers
%

\section{Introduction}

Asteroids preserve some of the earliest solid materials to have formed in the Solar System, yet their current location in the relatively narrow region of the asteroid belt does not reflect the large chemical, compositional, and isotopic diversity of meteorites, defying any simple \textit{in-situ} formation scenario. Dynamical models strongly suggest that these bodies formed across a broad range of heliocentric distances, from the terrestrial planet region ($\sim$1\,au) outward to the Kuiper Belt ($\sim$30\,~au), and were subsequently implanted into their present-day orbits by the gravitational influence and migration of the giant planets \citep{Morbidelli2005, Bottke2006, Levison2009, Walsh2011, Raymond2017a}. Recent meteorite analyses have further supported this view, highlighting an isotopic dichotomy between non-carbonaceous (NC) and carbonaceous chondrites (CC), implying distinct formation reservoirs \citep{Warren2011, Budde2016, Kruijer2017}. More subtle isotopic distinctions within the carbonaceous chondrite family, such as those between CI chondrites and other CC subtypes \citep{Hopp2022, Hellmann2023, Spitzer2024}, also suggest multiple reservoirs or formation epochs. Determining exactly where and when these bodies originated is thus critical for reconstructing the timing and mechanisms of giant planet formation, migration, and the delivery processes that shaped today's Solar System architecture. Asteroids, therefore, hold the key to unlocking the chronology and dynamical evolution of planet formation.

Spectroscopic observations have linked specific CC groups to large ($\gtrsim$100\,km) asteroids in the main belt \citep[e.g.,][]{Vernazza2016,DeMeo2022}, indicating that CM-like (Mighei-like) planetesimals are Ch/Cgh-type asteroids, whereas CI-like (Ivuna-like) bodies comprise B, C, Cb, and Cg-type asteroids \citep{DeMeo2013, DeMeo2022, Broz2024b, Anderson2025}. CI chondrites are particularly interesting in their lack of chondrules and CAIs, along with their specific Fe and Ti isotopic anomalies, leading some to suggest a trichotomy between the NC-CC-CI types \citep{Hopp2022, Hellmann2023, Spitzer2024}.

\begin{figure}[t!]
    \centering
    \includegraphics[width=\linewidth]{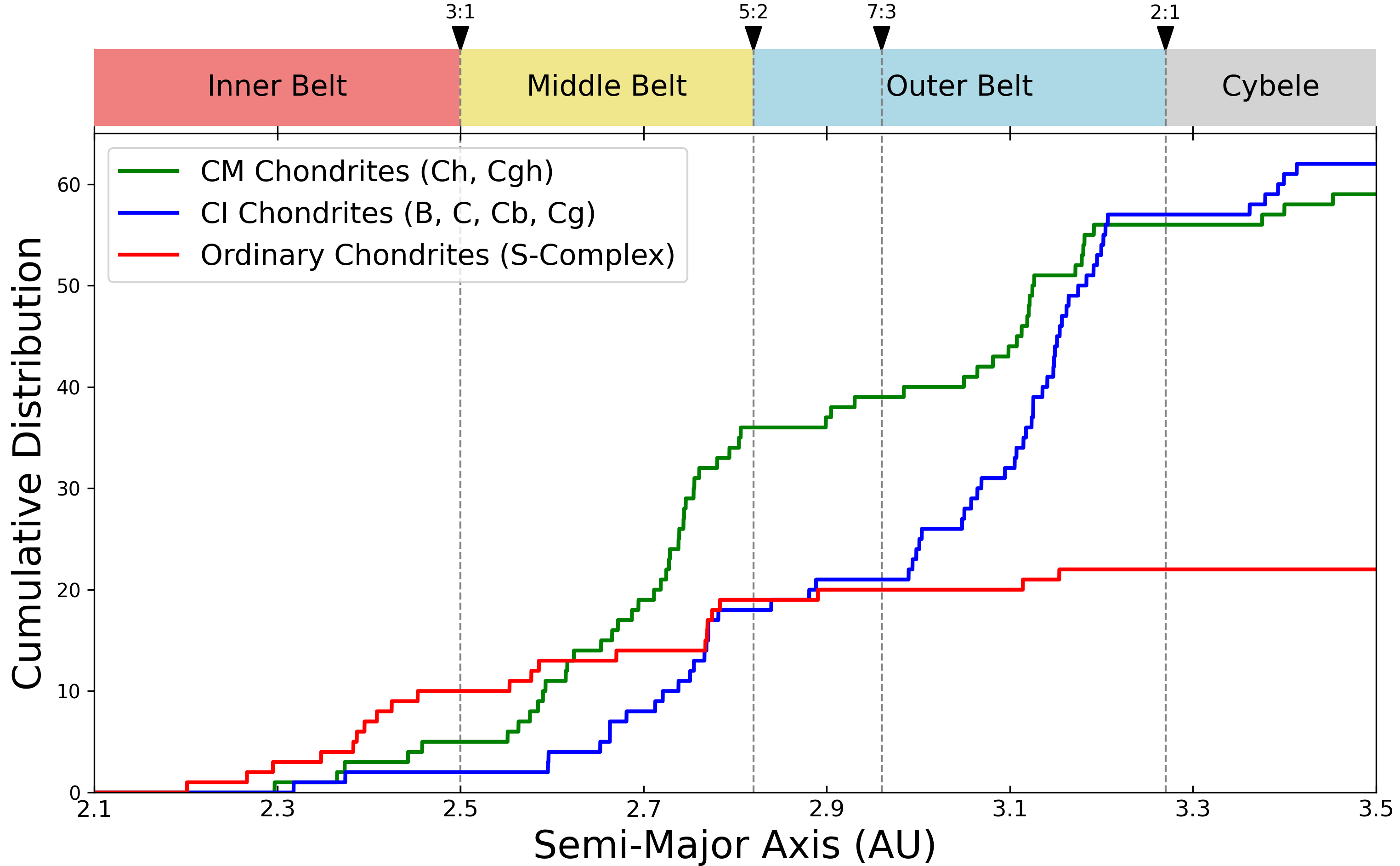}
    \caption{Cumulative distributions of the semi-major axis $N({\leq}a)$ of ${>}100$\,km asteroids classified as CM-like, CI/IDP-like (including B types with the geometric albedo of $p_V < 0.1$), and S-type. Vertical dotted lines mark major mean-motion resonances with Jupiter. CM-like objects exhibit an approximately symmetric distribution, whereas CI-like objects show a notably asymmetric distribution. Adapted from \cite{Anderson2025}.}
    \label{fig:fig1}
\end{figure}

Recently, it has been shown that the observed radial distributions of CM- and CI-like asteroids in the belt are distinct: CM-like objects exhibit an approximately symmetric distribution and are more abundant in the middle belt \citep{Rivkin2012, Anderson2025}, whereas CI/IDP-like objects show a notably asymmetric distribution and dominate in the outer belt \citep[see Fig.~\ref{fig:fig1} or][]{Anderson2025}. The latter is similar to the distribution of comet-like (P-type) asteroids \citep{DeMeo2013}, expected from late-stage, outer-disk delivery \citep{Nesvorn2018}. CI- and CM-like objects are also present in the belt at comparable overall levels (1.1:1 CI/CM). In our previous work \citep{Anderson2025}, we demonstrated that this dichotomy directly reflects two separate implantation events, reinforcing that CM and CI groups entered the asteroid belt from clearly distinct formation zones. We proposed that CM–like parent bodies formed in the Jupiter-Saturn region, in a pressure bump on the outer edge of the gap carved by Jupiter \citep[$<$10\,au;][]{Hellmann2023}. According to this scenario, Jupiter's early partial gap concentrated dust and pebbles at its edge, creating ideal conditions for local planetesimal formation. We demonstrated that a substantial fraction of these CM-like bodies were subsequently implanted into the asteroid belt, specifically during the rapid growth phase of Saturn, compatible with \citet{Raymond2017a}. By contrast, CI chondrites likely originated farther out, in the Uranus–Neptune zone \citep[$>$15\,au,][]{Hopp2022}, and were implanted into the belt at a distinctly later stage when these ice giants underwent their final accretion and migration phases \citep{Izidoro2015, Ribeiro2020, Nesvorny2024, Anderson2025}. This interpretation is consistent with their lack of chondrules, indicating that CIs directly accreted from primordial nebular dust that never experienced the intense heating events required to form chondrules. 

Independent constraints also link CI-like material to a more distant and likely later-forming reservoir: CIs share building-block properties with cometary materials such as IDPs \citep[100\% matrix, $\sim$0\% chondrules, $\sim$0\% CAIs;][]{Nakamura2023, Nakashima2023, Zega2025, Bradley1994}, and anhydrous areas in CI chondrites exhibit comet-like spectral behavior \citep{Brunetto2023}. Chronology is consistent with this picture, with CM ages clustering around $\sim$3–4\,Myr after CAIs \citep{Tanaka2024}, whereas IDPs, comet-like bodies, and trans-Neptunian objects are generally associated with formation after $\gtrsim$4–5\,Myr \citep{Neveu2019, Davidsson2016, Bierson2019}. Finally, the spectral similarity among P/D-type asteroids, comets, Jupiter Trojans, and Centaurs \citep{Vernazza2025}, together with models favoring delayed Uranus–Neptune formation \citep[$\sim$4–6\,Myr after CAIs;][]{Dodson2010, Helled2014}, naturally places CI-like delivery in a later evolutionary phase than CM-like implantation. Taken together, these constraints motivate treating the CM and CI source regions as radially separated reservoirs with different delivery windows, rather than neighboring sub-populations of a single carbonaceous reservoir. 

While the exact mechanism of chondrule formation remains uncertain, they require intense heating and high-pressure conditions to form, implying that their production was confined to regions closer to the young Sun and not in the distant ice-giant zones \citep{AliDib2025}. The relative lack of intermediate classes (CR, CK, etc.) in the belt raises the question of whether the CI formation region was geographically isolated. In short, to what extent was the CI reservoir truly "quarantined" from other CC groups, and could CM-like material be scattered outward? To our knowledge, this specific phase of outward planetesimals transfer scattered by Jupiter–Saturn growth before Uranus and Neptune began significant accretion and migration has not been explicitly quantified in previous dynamical studies.

Here, we explore the potential outward transport of ${>}100$-km-sized CM-like planetesimals initially formed on the outskirts of Jupiter’s gap to the Uranus-Neptune region. By quantifying the fraction of bodies that were scattered outward during Saturn’s growth, constrain the efficiency of outward transfer of CM-like planetesimals and assess whether any of this material can be retained on stable orbits. This provides a dynamical upper bound on the availability of chondrule-rich inner carbonaceous material to the Uranus–Neptune region and, by extension, to any populations sourced from that region. Section~\ref{sec:methods} describes the details of our dynamical simulations, including the respective disk model, planetesimal-disk interactions, and gravitational scattering by proto-planets. We present the main results in Section~\ref{sec:results}, focusing on the outward-transport efficiencies and timescales, followed by a discussion of their implications for the compositional diversity of the outer Solar System in Sec.~\ref{sec:discuss}. Finally, we summarize our conclusions in Section~\ref{sec:conclusions}.

\section{Methods}\label{sec:methods}

\citet{Raymond2017} demonstrated that, in a growing giant-planet system, planetesimals can be scattered to high eccentricities along curves of constant Tisserand parameter. Under sufficient gas drag, smaller bodies (1-10\,km) have their eccentricities rapidly damped and can survive on stable orbits in the outer disk. However, the gas density beyond Saturn’s orbit is too low to circularize large ($<$100\,km) planetesimals before they are ejected. Consequently, they are far more likely to be lost from the system. Building on their approach, we adopt a similar conceptual framework for outward scattering but focus on two new aspects: \textit{(i)} varying the physical conditions of the protoplanetary disk, \textit{(ii)} exploring a range of planet formation timescales in which Saturn and additional ice giant cores grow from $\sim$1\,$M_\oplus$ to their final masses, while allowing the growing cores to migrate inward, and (iii) employing different gas profiles to examine the impact of the density of the environment plays in the circularization of planetesimals' orbits. Moreover, this model is anchored in the idea that Saturn forms at the outer edge of Jupiter’s gap, consistent with pressure-bump formation scenarios \citep[e.g.,][]{Hellmann2023, Lau2024, Xu2024}.

\subsection{Initial state of the system}

As Jupiter's core grew to about $20 M_\oplus$, it began carving a gap in the protoplanetary disk, thereby inducing a local pressure bump that prevented pebbles from drifting inward \citep{Lambrechts2014}. This gap grew deeper as Jupiter accumulated mass, further isolating its vicinity from the surrounding disk \citep{Lin1986, Lee2002, Crida2017}. Eventually, Jupiter settled into a Type-II migration regime, slowing its radial motion and allowing pebbles to steadily accumulate at the outer edge of its gap \citep{Gonzalez2015}. This region experienced an elevated dust-to-gas ratio, creating favorable conditions for efficient planetesimal formation via cohesion or streaming instability. Consequently, a robust planetesimal reservoir is expected to have formed beyond Jupiter’s gap, while local material close to the planet’s orbit was largely depleted.

Previous works proposed that Saturn may have formed at or near this outer boundary \citep{Kobayashi2012, Lambrechts2014, Gonzalez2015, Lau2024}, which is consistent with evidence that Saturn once orbited closer to Jupiter than its present-day location \citep{Deienno2017}. As Saturn’s core grew, its increasing gravitational reach scattered these newly formed planetesimals inward, supplying solids to Jupiter’s circumplanetary disk (CPD), potentially contributing to the formation of the Galilean satellites \citep{Ronnet2018}. Many planetesimals were delivered toward the inner Solar System, either captured on stable orbits within the asteroid belt or flung farther inward, possibly aiding water delivery to the terrestrial planet region \citep{Meech2020}. The relative importance of each of these pathways depends strongly on gas-drag strength \citep{Raymond2017a}, which is itself influenced by both the disk density\textemdash thus its temporal evolution\textemdash and the sizes of the planetesimals, with smaller bodies experiencing more pronounced aerodynamic forces \citep{Raymond2022}. In general, outward scattering is difficult, since gas drag tends to shrink the orbit toward pericenter.

In the simulations presented here, we adopt the same disk model and growth timescales as in \citet{Anderson2025}, originally based on \citet{Ronnet2018}. While Jupiter is assumed to be fully formed, we grow all other planets from $1\,M_\oplus$ to their final mass through a simplified gas accretion prescription (Sec. \ref{sec:growth}). The dynamical evolution of planetesimals is traced under the influence of gas drag, gravitational interactions with proto-planets, and close encounters that can lead to scattering or capture (Sec. \ref{sec:sbd}).

\subsection{Giant planet growth and dynamics}\label{sec:growth}

We carried out orbital integrations using the open-source \texttt{REBOUND} library \citep{Rein2012} and its high-accuracy, adaptive-timestep IAS15 integrator. These simulations include two to three giant planets and a swarm of 10,000 massless test particles that serve as proxies for planetesimals. The simulations are evolved with a timestep of $10^{-2}/(2\pi)\,\text{yr}$. Test particles that collide with a planet or acquire hyperbolic trajectories are removed automatically to improve computational efficiency. Terrestrial planets are omitted from the model due to their comparatively small masses and more compact orbits, which minimally affect the dynamical evolution of the outer protoplanetary disk while requiring more integration steps and longer calculations.

\paragraph{Initial conditions.}
We begin with Jupiter placed at a heliocentric distance of 5.43\,au, consistent with post-disk evolution scenarios for the gas giants. Saturn’s core is positioned at 7.37\,au, near the outer edge of Jupiter’s gap and in a 3:2 resonance with Jupiter. In the three-planet scenario, an additional `ice-giant' core is initialized at 9.67\,au to form a 3:2, 3:2 multi-resonant chain based on \citet{Deienno2017}. These orbital configurations align with established constraints from the current planetary configuration, the Jupiter Trojans or the irregular satellites, and the secular tilt resonance within the cold Edgeworth-Kuiper belt observed today \citep{Nesvorn2018, Baguet2019}. While \citet{Ronnet2018} also tested scenarios in which Saturn is formed further out and migrates inward, we do not explore that alternative here, as it has been shown to be less consistent with Saturn’s final growth pattern, as well as with the effective delivery of solids into Jupiter’s circumplanetary disk and the asteroid belt. Instead, we focus on Saturn’s in-situ formation near the gap edge, which is consistent with a growing body of work arguing that giant planets commonly emerge in local pressure maxima \citep[e.g.,][]{Hellmann2023, Lau2024, Xu2024}. While these planetary cores are likely to migrate during their formation while gas remains before assuming the positions that will lead to the Nice model configuration \citet{Morbidelli2005, Nesvorn2018, Deienno2017}, it is the most stable point from which to start.

\paragraph{Growth.}
At the outset of the simulations, Jupiter is taken to have its present-day mass (${\approx}300\,M_\oplus$). The other planetary cores begin at $1\,M_\oplus$ and increase in mass over a prescribed timescale $\tau_\text{growth}$, with Saturn reaching approximately $100\,M_\oplus$ and the additional ice-giant core growing to roughly $15\,M_\oplus$, comparable to the present-day mass of Uranus. This simplified growth prescription is adapted from \citet{Ronnet2018} and models the planet’s mass evolution as described by:

\begin{equation}
M_\text{planet}(t) = M_i + \Delta M \big(1-{\rm e}^{-t/\tau_\text{growth}}\big)\,,
\end{equation}

\noindent where $M_i$ is the initial core mass and $\Delta M$ is the difference between each planet's initial core mass and the final mass. This approach approximates standard core-accretion models, in which a gaseous envelope gradually contracts, transitions to runaway accretion, and then tapers off as the planet begins opening a gap \citep{Pollack1996}. We explore the impact of the formation timescale  with $\tau_\text{growth}$ set to $1 \times 10^5$, $5 \times 10^5$, and $1 \times 10^6$. Although more detailed models are needed to capture all aspects of giant planet growth, a precise reconstruction is beyond the scope of this study.

Although the exact timing for the emergence of an ice-giant embryo relative to Saturn’s formation is not fully constrained, we initialize both cores together to explore how an additional growing planet might affect the outward scattering of CM-like bodies. This choice is partly motivated by computational efficiency, as a second core emerging later would require restarting the simulation at a different epoch, and by the recognized possibility of multiple protoplanetary embryos coexisting in resonance chains \citep{Guilera2011,Lau2024}. In practice, setting the ice-giant core to grow in parallel with Saturn captures a range of plausible scenarios, whether this additional planet evolves into Uranus, Neptune, or the hypothetical fifth giant planet (so-called "Ice~1") postulated by the Nice Model \citep{Nesvorny2018, Deienno2017}. Although simplified, this approach allows us to bracket the influence of a simultaneous second core on scattering dynamics while avoiding uncertainties about the precise timing or radial location of a late-emerging ice giant.

\paragraph{Migration and damping.}

We account for the migration and damping of the growing planets due to gas-disk interactions, adopting fictitious accelerations as described by \citet{Ronnet2018}, based on \citet{Cresswell2008}. Specifically, we employed the following acceleration terms for migration, eccentricity, and inclination damping:

\begin{equation}
\mathbf{a}_\text{mig} = -\frac{\mathbf{v}}{\tau_\text{mig}}\,,
\end{equation}

\begin{equation}
\mathbf{a}_\text{e} = -2\frac{(\mathbf{v}\cdot\mathbf{r})\mathbf{r}}{r^2\tau_\text{e}}\,,
\end{equation}

\begin{equation}
\mathbf{a}_\text{i} = -\frac{v_z}{\tau_i}\mathbf{k}\,,
\end{equation}

\noindent where $\mathbf{v}$ is the velocity vector of the planet, $\mathbf{r}$ is its position vector, $\mathbf{k}$ is the unit vector in the $z$-direction, and $r$ is the heliocentric distance.

The prescription of the migration timescale is given by:

\begin{equation} 
\begin{split}
\tau_\text{mig} = & \frac{2t_{\text{wave}}}{2.7 + 1.1\beta}\bigg(\frac{H}{r}\bigg)^{-2}\Bigg[ P(e) + \frac{P(e)}{|P(e)|} \times \\
& \Bigg(0.070\bigg(\frac{i}{h}\bigg) + 0/085\bigg(\frac{i}{h}\bigg)^4 - 0.080\bigg(\frac{e}{h}\bigg)\bigg(\frac{i}{h}\bigg)^2\Bigg)  \Bigg], 
\end{split}
\end{equation}

\begin{equation} 
P(e) = \frac{1 + (\frac{e}{2.25 h})^{1.2} + (\frac{e}{2.84 h})^{6}}{1 - (\frac{e}{2.02 h})^{4}}
\end{equation}

\noindent where $i$ is the orbital inclination, $e$ the eccentricity, $h=H/r$ is the disk’s aspect ratio, $M_*$ denotes the stellar mass, $m_\text{p}$ the planet’s mass, $a_\text{p}$ its semi-major axis, and $\Omega_\text{p}$ the Keplerian frequency. Although the sign dependence of $P(e)$ allows torque reversal at sufficiently high eccentricities, our planets never enter that regime, and no outward migration arises. We set $\beta$ to 1.5 to be consistent across all gas profiles.

The damping timescale is given by: 

\begin{equation}
t_{\text{wave}} = \frac{M_*}{m_\text{p}}\frac{M_*}{\Sigma_\text{p}a_\text{p}^2}h^4\Omega_\text{p}^{-1} \,.
\end{equation}
The eccentricity damping time $\tau_e$ and inclination damping time $\tau_i$ are then calculated via:
\begin{equation} 
\tau_e = \frac{t_{\text{wave}}}{0.780},\Bigg[ 1 - 0.14\Big(\frac{e}{h}\Big)^2 + 0.06\Big(\frac{i}{h}\Big)^3 + 0.18\Big(\frac{e}{h}\Big)^2\Big(\frac{i}{h}\Big)\Bigg], 
\end{equation}
\begin{equation}
\tau_i = \frac{t_{\text{wave}}}{0.544}\times\bigg[ 1 - 0.33\bigg(\frac{i}{h}\bigg)^2\! + 0.24\bigg(\frac{i}{h}\bigg)^3\! + 0.14\bigg(\frac{e}{h}\bigg)^2\bigg(\frac{i}{h}\bigg) \bigg]   \,.
\end{equation}

We neglect higher-order torques (e.g., heating torques from rapidly accreting protoplanets \citealp{Benitez2015}; eccentric forcing from nonzero-eccentricity gaps \citealp{Fendyke2014}; or hot-trail effects \citealp{Eklund2017, Chrenko2017, Cornejo2023}) since our primary aim is to track the implantation of small bodies rather than to optimize an exact migration prescription \citep{Paardekooper2010, Paardekooper2011}.

This simplified disk model and set of growth prescriptions allowed us to study the implantation of planetesimals in the inner solar system \citep[e.g.,][]{Ronnet2018, Anderson2025}, and will now allow us to focus on whether and how Saturn’s formation efficiently scatters CM-like planetesimals outward, potentially delivering them to the Uranus–Neptune zone.

\subsection{Small body dynamics}\label{sec:sbd}

Observations and dynamical studies suggest that many small bodies populating the outer Solar System originated between Jupiter’s orbit and approximately 30\,au \citep{Tsiganis2005, Gomes2003, Levison2008, Kaib2008}. Here, we focus on the region immediately beyond Jupiter, where the CM-like bodies in this scenario are hypothesized to form. We randomly initialize 10,000 massless test particles representing planetesimals uniformly in semi-major axis in the $7$$<$$a$$<$$10$\,au range (two-planet simulation) and the $7$$<$$a$$<$$11$\,au range (three-planet simulation). This simplified approach does not attempt to reproduce any particular radial distribution of solids, which is poorly constrained and likely consisted of multiple pressure bumps or ring-like substructures \citep{Izidoro2022}. Instead, it ensures that each radial slice in that zone is adequately sampled so we can identify how far outward these bodies might travel in response to planetary perturbations. Varying the precise radial or size distribution would not drastically change the likelihood of their scattering since planetesimal–planetesimal interactions are negligible in comparison to the gravitational and aerodynamic forces exerted by the planets and the gaseous disk \citep{Raymond2017}.

We restrict the test particles to bodies of 100\,km in diameter. As shown by \citet{Raymond2017}, smaller ($\sim$1--10\,km) planetesimals are more likely to be captured onto stable orbits by aerodynamic drag once scattered outward, but they generally do not travel as far from their initial orbits ($\sim 5$\,au outward). In contrast, larger bodies can, in principle, be scattered to greater heliocentric distances, although the weaker drag force they experience makes it harder for them to circularize and remain in the system. By concentrating on these 100\,km objects, we effectively probe a `worst-case' limit for outward migration: if these 100\,km objects fail to be implanted in the Uranus–Neptune zone, we can infer that CM-like material of similar or larger sizes is unlikely to have contributed significantly to this region.

\paragraph{Aerodynamic drag.}
We investigate how the aerodynamic drag influences each planetesimal using the methods described by \citet{Ronnet2017} by applying the following acceleration term:

\begin{equation}
\mathbf{a}_\text{drag} = -\frac{1}{t_\text{s}}(\mathbf{v}-\mathbf{v}_\text{g})\,,
\end{equation}

\noindent where $\mathbf{v}$ is the planetesimal’s velocity, $\mathbf{v}_\text{g}$ is the local gas velocity, and $t_\text{s}$ is the stopping time. Following \citet{Perets2011} and \citet{Guillot2014}, we calculate $t_\text{s}$ for 100\,km-sized bodies of density \(\rho_\text{s} = 1\,\text{g}\,\text{cm}^{-3}\) as:

\begin{equation}
t_\text{s} = \left(\frac{\rho_\text{g} v_{\text{th}}}{\rho_\text{s} R_\text{s}}\text{min}\left[1, \frac{3}{8} \frac{v_{\text{rel}}}{v_{\text{th}}}C_\text{D}\right]\right)^{\!\!-1}\,,
\end{equation}

\noindent where $R_\text{s}$ is the size of the solids, here set to 100\,km, and $\rho_\text{s} = 1\,\text{g}\,\text{cm}^{-3}$ their density. The gas volumetric density $\rho_\text{g}$ is obtained from the surface density $\Sigma_\text{g}$ by assuming hydrostatic equilibrium in the vertical direction, with a disk aspect ratio of $h \equiv H/r = 0.05$ \citep{Cresswell2008}. The relative velocity between the gas and the planetesimal is $v_{\text{rel}}$, while the gas thermal velocity is given by $v_{\text{th}} = \sqrt{8/\pi}c_\text{s}$, where $c_\text{s}$ is the isothermal sound speed. In this context, the isothermal sound speed is defined as $c_\text{s} = \Omega H$, being $\Omega$ the Keplerian orbital frequency, expressed as $\Omega = \sqrt{GM_\odot/r^3}$. The dimensionless drag coefficient $C_\text{D}$ is computed as a function of the Reynolds number ${\rm Re}$ of the flow around the planetesimal \citep{Perets2011}:

\begin{equation}
C_\text{D} = \frac{24}{{\rm Re}}(1+0.27{\rm Re})^{0.43} + 0.47 \left(1 - {\rm e}^{-0.04{\rm Re}^{0.38}}\right)\,,
\end{equation}
\begin{equation}
{\rm Re} \equiv \frac{4R_\text{s}v_{\text{rel}}}{c_\text{g}l_\text{g}}\,.
\end{equation}

\noindent Using the mean free path of the gas $l_\text{g}$ from \citet{Supulver2000} and the size of 100\,km, we find a value of $C_\text{D} = 0.44$, which is then fixed as a constant to save computational time.


\subsection{Gas profiles}

\begin{figure}
    \centering
    \includegraphics[width=\columnwidth]{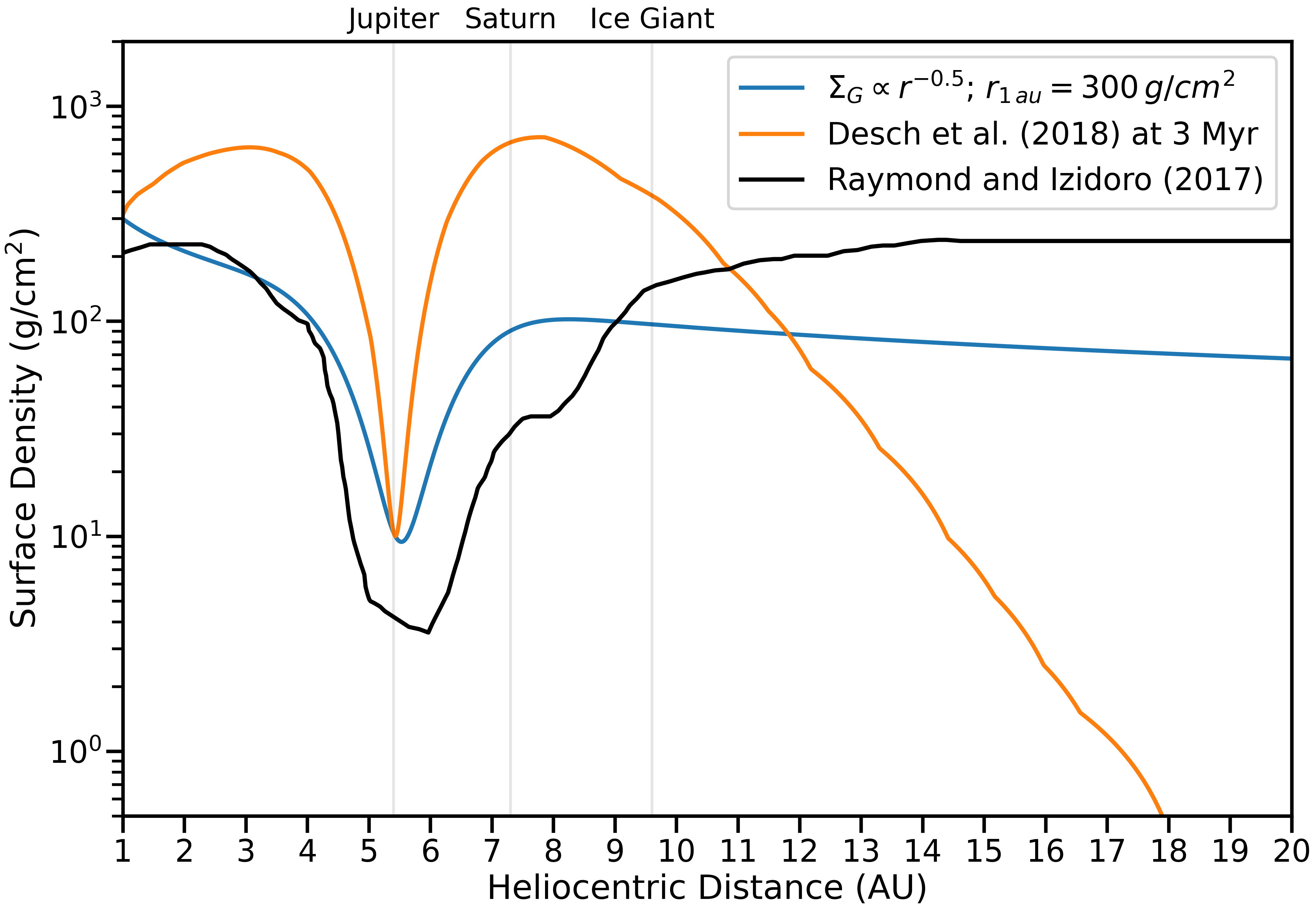}
    \caption{Comparison of the gas profiles used in the various simulations. The initial orbits of Jupiter, Saturn's core, and the additional ice giant core are indicated by grey vertical lines. Beyond 10\,au, the profiles can be seen as Flat \citep[Black,][]{Raymond2017a}, Sloped \citep[Blue,][]{Cresswell2008}, and with a Sharp Drop-off \citep[Orange,][]{Desch2018}.}
    \label{fig:SigmaG}
\end{figure}

Accurately defining the disk’s gas distribution is notoriously challenging due to the complex interplay of processes such as radiative transfer, dust condensation, viscous heating, and planet-disk interactions \citep[e.g.,][]{Nelson2000, Bitsch2015, Johansen2015}. As a result, no single disk model has emerged as definitive. In our previous study \citet{Anderson2025}, we showed that subtle differences in adopted gas profiles can have a profound impact on how, when, and where planetesimals are implanted into the inner Solar System. Here, we extend that analysis to the outer region beyond Saturn, where the physical conditions are even less certain. We employ the same three different disk models in order to capture the range of outcomes for planetesimal scattering and planetary embryo migration. A comparison of all these gas surface density profiles is shown in Fig.~\ref{fig:SigmaG}. Since the gas profile of the outer disk is even less constrained than the inner system, we can describe them as follows:

\begin{table*}[t!]
\centering
\caption{Planetesimal distribution for our two-planet scenario at $t=300\,\mathrm{kyr}$ for each gas profile and Saturn growth timescale, expressed as percentages of the initial $\sim$10{,}000 test particles.}
\label{tab:2Pl}
\begin{tabular}{lccccccccc}
\hline\hline
Gas  & $\tau_\mathrm{G}$ & Removed 
& $a$$<$$5$\,au &$5$$<$$a$$<$$10$\,au & $10$$<$$a$$<$$15$\,au & $10$$<$$a$$<$$15$\,au & $15$$<$$a$$<20$\,au & $15$$<$$a$$<20$\,au & $a$$>$$20$\,au \\ 
Profile &  &  & $e<0.4$ & $e<0.4$ & $e<1$ & $e<0.4$& $e<1$ & $e<0.4$ & $e<1$ \\ 
\hline
Flat               & $1\times10^5$ & 31.60 & 4.50 &	47.10 &	14.30 & 14.15 &	                    2.02 & 1.45 & 0.48 \\
                   & $5\times10^5$ & 21.45 & 3.58 & 58.77 & 13.22 & 12.74 & 2.52 & 1.74 & 0.45 \\
                   & $1\times10^6$ & 17.12	& 3.42 & 62.84 & 13.48 & 12.89 & 2.57 & 1.64 & 0.87 \\
                   
Sloped             & $1\times10^5$ & 33.20 & 5.21 &	51.08 &	8.85 & 8.45 &                        1.05 & 0.14 & 0.48 \\
                   & $5\times10^5$ & 26.96 & 4.64 & 59.48 & 6.91 &	6.11 & 1.34 &	0.08 & 0.67  \\
                   & $1\times10^6$ & 21.96	& 3.81 & 64.80 & 7.27 &	5.97 & 1.30 & 0.04 & 0.87 \\

Sharp              & $1\times10^5$ & 19.94 & 5.94 & 69.86 & 4.00 & 3.81 &                        0.09 & 0.00  & 0.17 \\
drop-off           & $5\times10^5$ & 12.51	& 4.50 & 81.91 & 1.64 & 1.44 &                       0.06 &  0.00 & 0.09 \\
                   & $1\times10^6$ &  9.39 & 3.63 & 85.49 & 1.34 & 1.02 & 0.07 & 0.00 & 0.08 \\
\hline

\end{tabular}

\end{table*}

\paragraph{Sloped profile.} A commonly used 1-D model following a power-law decrease in gas surface density, typically with $\Sigma_\text{g} \propto r^{-0.5}$ normalized to $\Sigma_\text{g}=300\,\mathrm{g}\,\mathrm{cm}^{-2}$ at 1\,au \citep{Ronnet2018} along with the aspect ratio $h = 0.05$ \citep{Cresswell2008}. Jupiter’s partial gap is introduced by reducing the surface density by a factor of $10^{-1}$ near the planet’s orbit \citep[see also][]{Bitsch2015}, mimicking the deepening of the gap as Jupiter is already close to its final mass. 
    
\paragraph{Flat profile.} \citet{Raymond2017a} assumed the surface density profile from \citet{Morbidelli2007}, starting after Jupiter and Saturn were both formed. The starting disk profile has a $\Sigma_\text{g}$ of 200\,g\,cm$^{-2}$ at 1\,au, a gap opened by Jupiter at 5.4\,au, and a partial gap opened around Saturn at 7\,au. While they decreased the surface density uniformly in radius, on a $2 \times 10^5$\,yr exponential timescale, until it was entirely removed by $2 \times 10^6$\, yr, we chose to examine their starting disk profile, as the gap opened by Jupiter and Saturn was already much wider than in the other profiles. The model is practically flat beyond 10\,au, and allows us to test how a higher residual gas density may allow some larger solids to damp their eccentricities enough to survive outward scattering, although it also influences the migration speed of the growing embryos.
   
\paragraph{Sharp drop-off profile.} \citet{Desch2018} considered a self-similar disk model \citep{Hartmann1998} that evolves in time through viscous diffusion. It features a higher initial surface density at small radii and a steep slope beyond 8\,au. While they examined a gap location of $\sim$3\,au for Jupiter, we apply a similar prescription to a gap at 5.4\,au, consistent with our Saturn-growth scenario and the notion that Jupiter opened its gap closer to its current orbit. In this model, the precipitous decline in $\Sigma_\text{g}$ beyond Saturn reduces the efficacy of aerodynamic damping for large planetesimals, potentially limiting their ability to remain in stable orbits at high eccentricities. We select their 3 Myr gas profile, as it is an intermediary profile, and should approximate the time after Jupiter has formed.
\vskip\baselineskip

The disk viscosity is parameterized by the classical $\alpha$ prescription \citep{Shakura1973}, with a typical $\alpha \approx 2\times 10^{-3}$. We do not investigate this parameter as we found it has a modest effect on our previous simulations. More details can be found in our \citet{Anderson2025} paper as well as the respective papers that originated the individual profiles.

By comparing these three scenarios, we aim to bracket the potential range of outcomes for how far CM-like planetesimals might be transported outward and whether they can be implanted in the Uranus-Neptune range. Given the uncertainties inherent to protoplanetary disk evolution at large radii, these models should provide a robust test of how the outer disk profile shapes the fate of planetesimals initially located near Jupiter’s gap edge.

\begin{figure*}[p]
\centering
    \includegraphics[width= \linewidth]{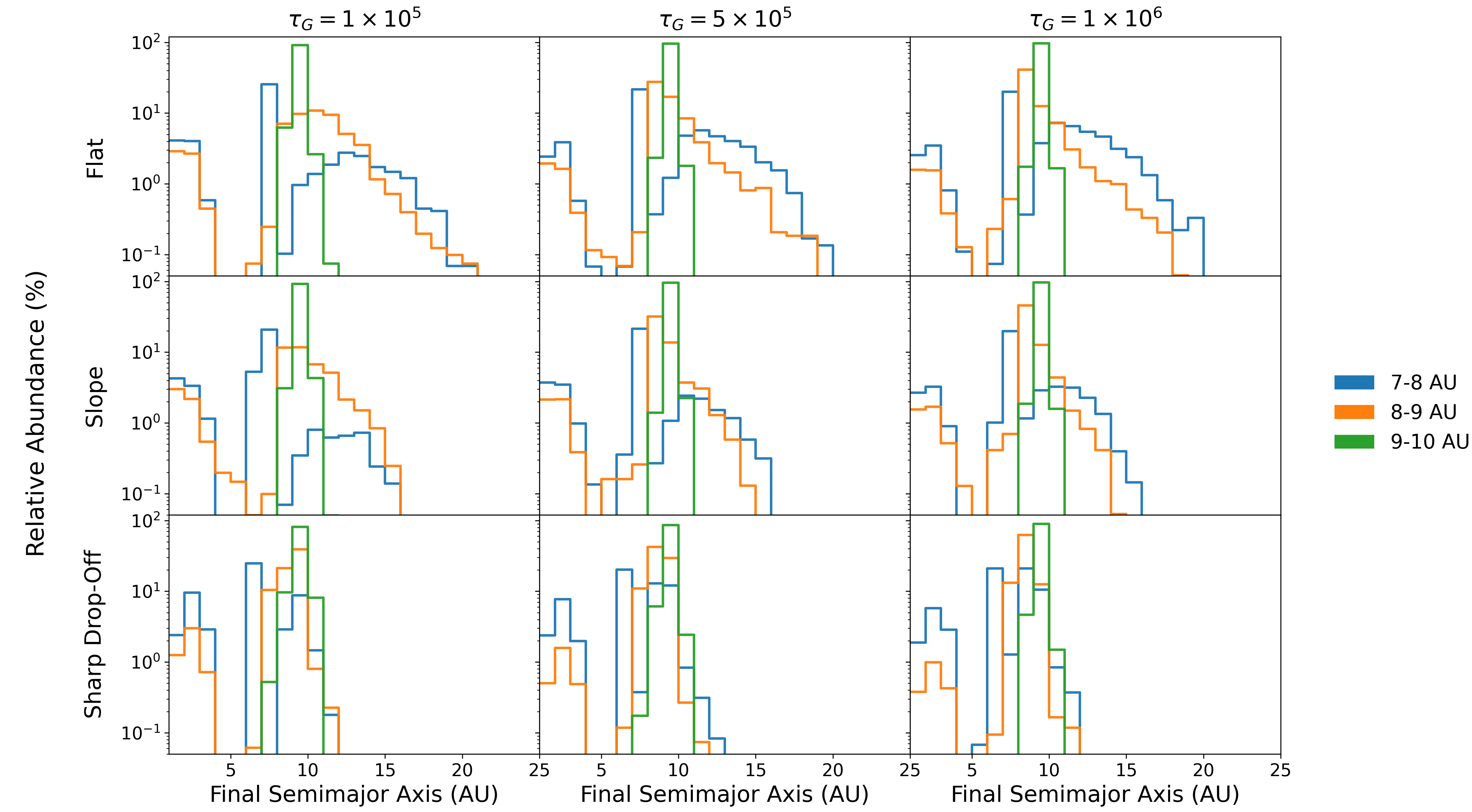}
    \caption{Fraction (\%) of planetesimals from each initial formation zone that end up within 1-au-wide semi-major-axis bins at $t = 300,\mathrm{kyr}$ for the two-planet model (log scale). We consider implanted objects as those with $e<0.4$. In all cases, most bodies formed between 9 and 10\,au remain near their birth locations, reflecting a dynamically quiet zone largely unaffected by Saturn’s growth. The prominent blue feature near $\sim$7\,au corresponds to particles trapped in Saturn’s co-orbital region.}
    \label{fig:2P_Imp}

    \includegraphics[width=0.9\linewidth]{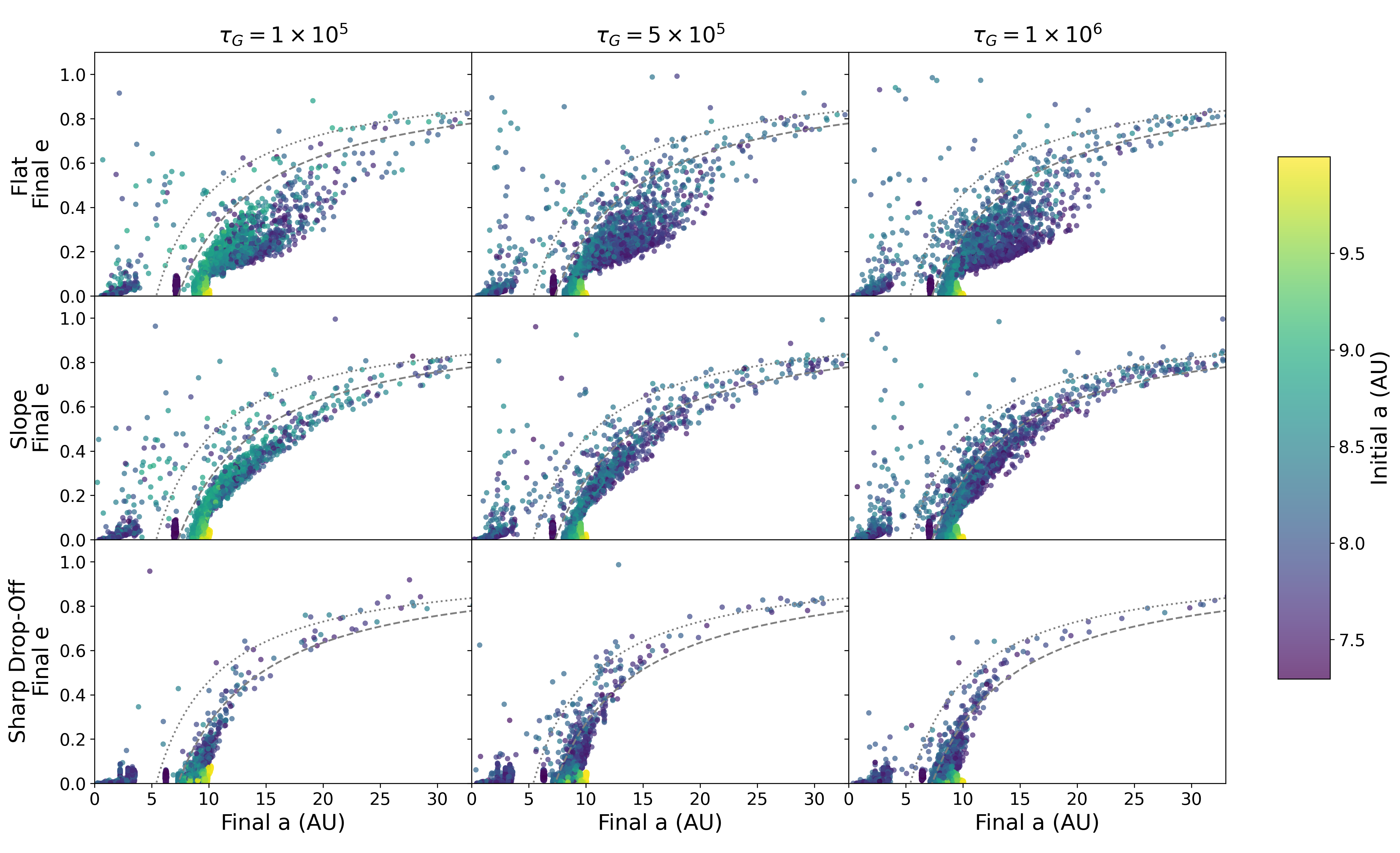}
    \caption{Eccentricity of remaining planetesimals as a function of semi-major axis at $t = 300\,\mathrm{kyr}$ for each gas profile and Saturn growth timescale for our two-planet model. The dotted line represents Jupiter-crossing orbits,  the dashed line represents Saturn-crossing orbits. The color represents the planetesimals' formation location, with cooler colors representing earlier loss and yellow for those that remain. The more gas in the simulation, the more easily the planetesimals are circularized (lower final $e$). The cluster of purple points near $\sim$7 au represents planetesimals trapped near Saturn’s orbit; these bodies are dynamically influenced by the planet and are likely to be absorbed into its circumplanetary disk over time.}
    \label{fig:2P_AE}
\end{figure*}


\section{Results}\label{sec:results}

\subsection{Two planet scenario}

In all of the two-planet runs, type-I migration drives a modest inward migration of Saturn, though this inward shift amounts to only a few tenths of an astronomical unit by the end of the simulations. However, under the \citet{Desch2018}-inspired disk profile, where the surface density remains higher $<$10\,au, both Jupiter and Saturn migrate more substantially, essentially because the deeper reservoir of gas exerts a stronger net torque on them.

As expected, the gravitational influences of Jupiter and the growing Saturn excite the eccentricities of nearby planetesimals, leading to scattering both inward (toward the asteroid belt and terrestrial region) and outward (toward semi-major axes beyond 10\,au). Right away, Saturn’s core scatters many planetesimals that happen to lie within roughly $\pm1$\,au of its orbit, depending on initial configuration (Fig.~\ref{fig:2P_Imp}). These objects follow characteristic `wings' in ($a,e$)-space (Fig.~\ref{fig:2P_AE}), corresponding to curves of constant Tisserand parameter \citep{Levison1997}. 

While we ran the simulations out to $1\times 10^6$ years, we note that the most significant scattering and planetesimal removal occurred within the first $2.5\times 10^5$ years. By this point, the bulk of strongly perturbed bodies have either collided with Saturn, become trapped within Jupiter’s circumplanetary region, or been ejected from the system entirely. Subsequent evolution, beyond a few hundred thousand years, largely results in the slow inward migration of Saturn without any appreciable new scattering events. Since many simulations also lost all of their planetesimals before reaching the 1\,Myr mark, we focus on outputs after $2.5\times 10^5$ years, where the fate of individual planetesimals is already well established and further integration offers diminishing returns. A direct comparison of the final distributions across the different gas profiles at $t=300$\,kyr is provided in Table~\ref{tab:2Pl}.

\paragraph{Initial position.}
All planetesimals that have been completely removed from the system originated in the $7.2-8.6$\,au zone (Fig.~\ref{fig:3P_Init}). These represent a quarter to a third of all initialized planetesimals. Their fates vary: some are accreted onto Saturn, some are driven onto orbits that deliver them to Jupiter’s circumplanetary disk, and others are ejected from the Solar System. Any planetesimals formed in this range that remain after this initial scattering event typically settle onto stable orbits either in the inner Solar System, becoming implanted in the asteroid belt or even the terrestrial formation zone, or occupy more distant, excited orbits, in the 10-20\,au range.

Planetesimals having formed beyond $\sim$8.6\,au generally retain semi-major axes close to their initial locations but develop higher eccentricities. Although Saturn stirs these outer objects, it does not scatter them strongly inward or outward on these timescales. Meanwhile, any bodies initiated inside $\sim$7.5\,au are quickly caught co-orbiting Saturn (Fig.~\ref{fig:2P_AE}) on trajectories that would ultimately deliver them into Saturn’s Hill sphere and feed the circumplanetary region. In these simulations, which do not include an explicit circumplanetary disk, such bodies are treated as ‘absorbed’ once they collide with the planetary surface. In principle, a more detailed prescription that includes a circumplanetary disk \citep[as performed in][]{Ronnet2018} could track whether some fraction remains in bound orbits around Saturn, potentially contributing to satellite formation.

\paragraph{Growth timescale.}
An additional parameter we tested is the growth timescale $\tau_\text{growth}$ for Saturn. Unsurprisingly, faster core accretion means that Saturn reaches a higher mass sooner, which tends to enhance its migration and early scattering of nearby planetesimals. In simulations with a faster $\tau_\text{growth}$, a third of the test particles are removed, while a slower $\tau_\text{growth}$ allows us to retain three-quarters. Yet these results show that the final distribution of surviving objects is almost indistinguishable across different $\tau_\text{growth}$ values. The system simply evolves more quickly when Saturn is growing faster but ultimately settles into a similar end state, with the same fraction of planetesimals stranded in the inner Solar System or scattered outward. In particular, for growth timescales ranging from $\tau_\text{growth} = 1\times 10^5, ~5\times 10^5$, and $1\times 10^6$ years, the overall clearing zone and the fraction of removed bodies remain roughly the same. This implies that how quickly Saturn reaches its final mass has little bearing on which planetesimals survive, aside from minor variations in their removal timescales.

\begin{figure*}[p]
    \centering
    \includegraphics[width=\linewidth]{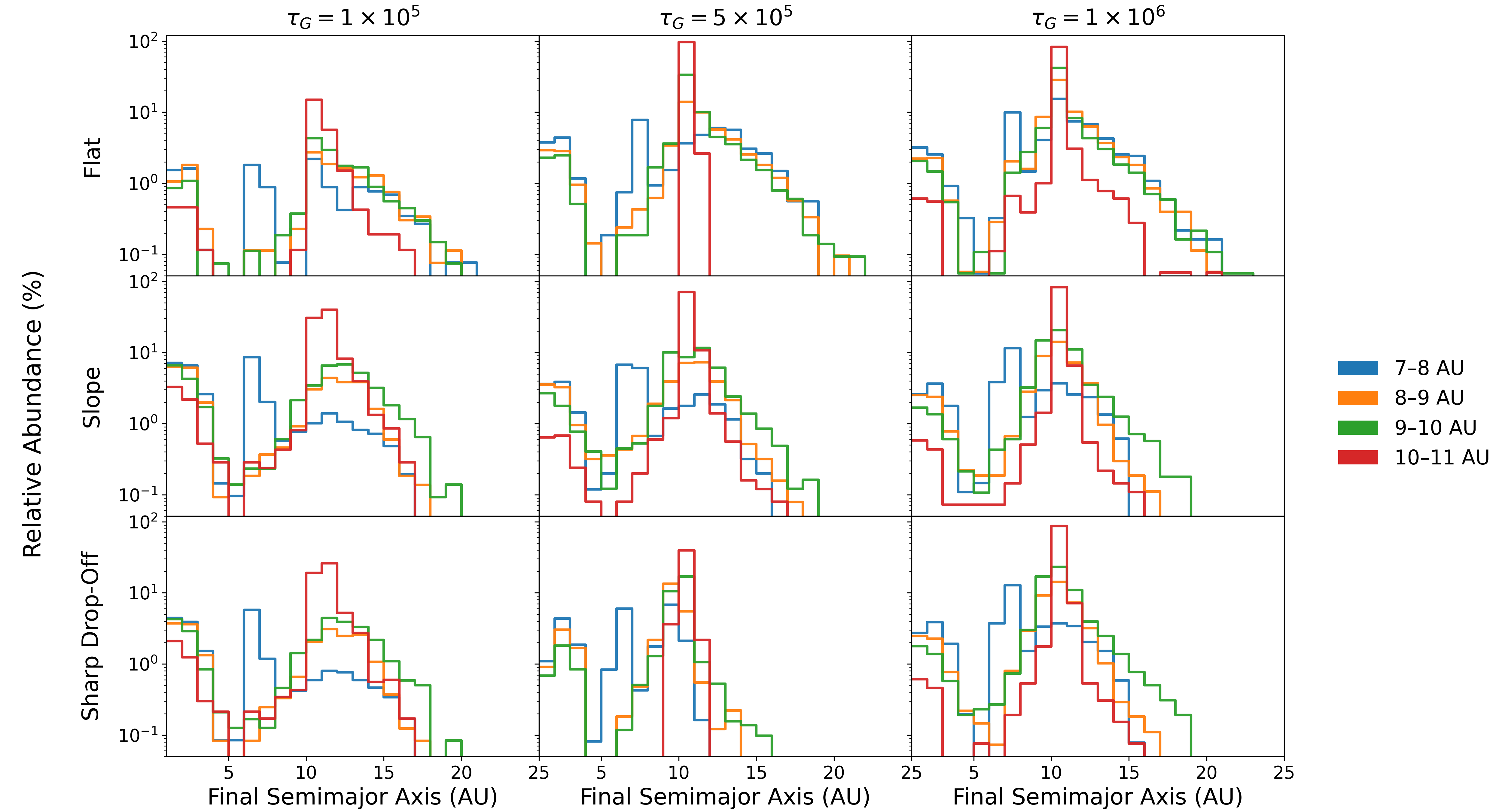}
    \caption{Same as Fig.~\ref{fig:2P_Imp}, but for the three-planet model. We consider implanted objects as those with $e<0.4$. Faster planetary growth clears the local planetesimal population. Objects that formed beyond 10 au are less likely to interact with the forming planetesimals.}
    \label{fig:3P_Imp}
    \vspace{1cm}
    \includegraphics[width=\linewidth]{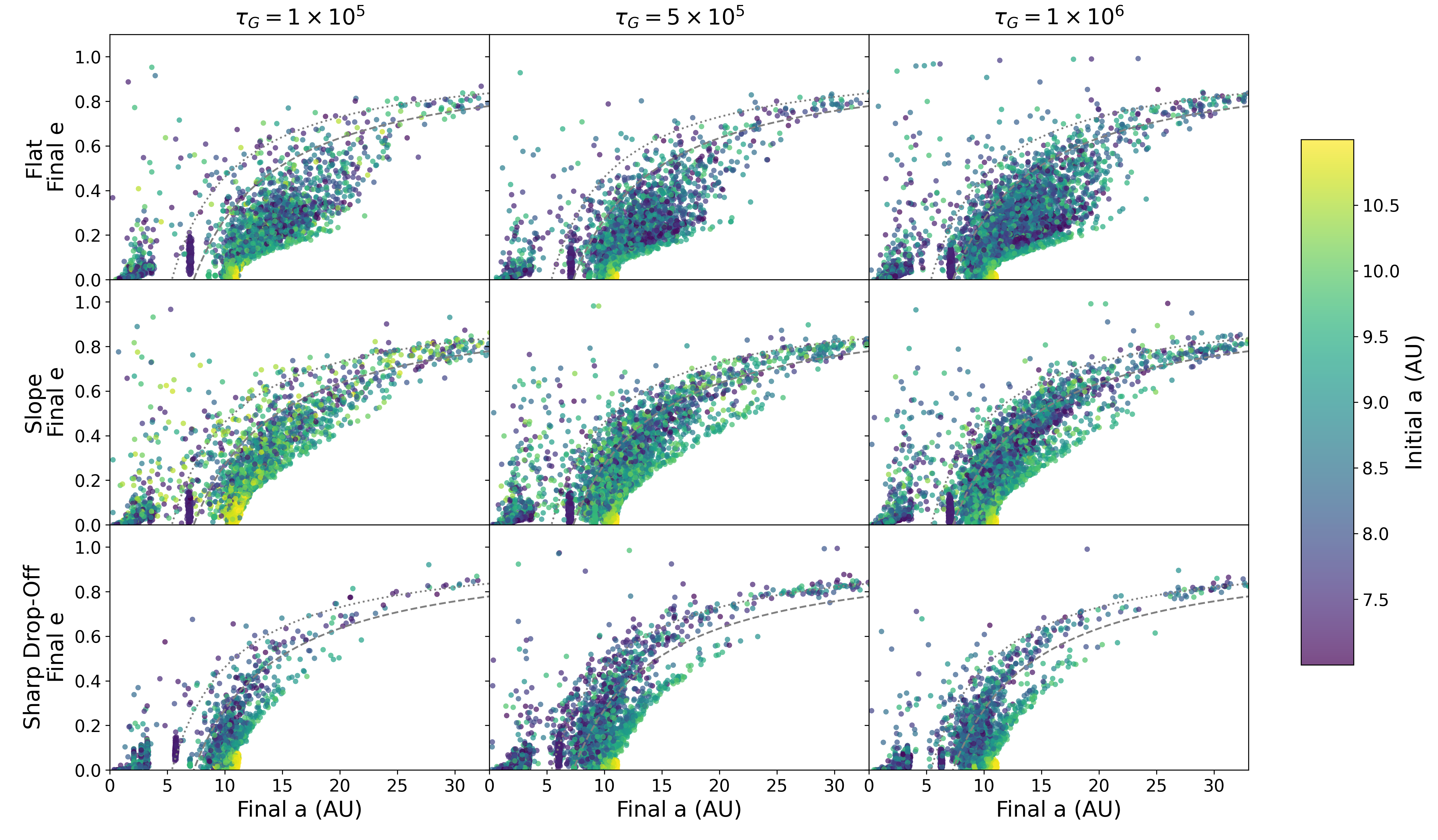}
    \caption{Same as Fig.~\ref{fig:2P_AE}, but for the three-planet model. The dotted line represents Jupiter-crossing orbits, and the dashed line represents Saturn-crossing orbits.  The color represents the planetesimals' formation location, with cooler colors representing earlier loss and yellow for those that remain. The more gas in the simulation, the more easily the planetesimals are circularized.}
    \label{fig:3P_AE}
\end{figure*}

Saturn’s orbital evolution remained modest, consistent with the findings of \citet{Raymond2017a}, thanks to its resonance lock with Jupiter and the partial gap carved by the two planets. Planetesimals within roughly 1\,au on either side of Saturn’s orbit were rapidly scattered, following constant-Tisserand pathways and eventually reaching either the inner Solar System or high-eccentricity orbits in the 10-20\,au region. However, since neither the surface density nor the gas drag was reduced over time in our simulations, the abundant gas supply acted to circularize many orbits rather than eject them outright, allowing a fraction of larger bodies to survive beyond 20\,au. Interestingly, we also find that scenarios with lower gas density beyond $\sim10$\,au can allow scattered planetesimals to reach even larger semi-major axes, yet these same runs yield fewer objects with $e<0.4$ on stable orbits. Strong aerodynamic damping is needed to retain them at such distances: a higher disk gas density increases the fraction of bodies retained on orbits beyond 10\,au, since aerodynamic drag can circularize orbits before they are fully ejected.

\begin{table*}[t!] 

\caption{Planetesimal distribution for our three-planet scenario at $t=300~\mathrm{kyr}$ for each gas profile and Saturn growth timescale, expressed as percentages of the initial $\sim$10{,}000 test particles.}
\label{tab:3Pl}
\centering
\begin{tabular}{lccccccccc}
\hline\hline
Gas  & $\tau_\mathrm{G}$ & Removed 
& $a$$<$$5$\,au &$5$$<$$a$$<$$10$\,au & $10$$<$$a$$<$$15$\,au & $10$$<$$a$$<$$15$\,au & $15$$<$$a$$<20$\,au & $15$$<$$a$$<20$\,au & $a$$>$$20$\,au \\ 
Profile &  &  & $e<0.4$ & $e<0.4$ & $e<1$ & $e<0.4$& $e<1$ & $e<0.4$ & $e<1$ \\ 
\hline
Flat               & $1\times10^5$ & 41.83 &  7.58 &  2.75 & 41.40 & 40.83 &                    5.05 & 3.78 & 1.39 \\
                   & $5\times10^5$ & 38.63 &  5.59 &  5.61 & 54.40 & 53.28 & 4.78 & 3.10 & 1.19 \\
                   & $1\times10^6$ & 22.65 &  4.47 &  9.35 & 57.27 & 56.01 & 5.07 & 3.19 & 1.18 \\
Sloped             & $1\times10^5$ & 54.95 &  9.84 &  3.99 & 24.77 & 23.35&                     3.91 & 1.20 & 2.54 \\
                   & $5\times10^5$ & 38.11 &  7.02 & 10.66 & 38.66 & 35.86 & 3.47 & 0.55 & 2.08 \\
                   & $1\times10^6$ & 30.44	&  5.45 & 14.26 & 44.54 & 41.71 & 3.58 & 0.54 & 1.72 \\
Sharp              & $1\times10^5$ & 32.55 & 10.54 & 14.04 & 42.19 & 41.52 &                    0.41 & 0.03 &  0.28 \\
drop-off           & $5\times10^5$ & 23.64 &  9.42 & 26.67 & 39.37 & 38.51 &                    0.51 &  0.05 &   0.38 \\
                   & $1\times10^6$ & 12.52	&  7.61 & 33.25 & 38.57 & 37.70 & 0.57 & 0.08 & 1.48 \\
\hline
\end{tabular}
\end{table*}

Because gas drag damps eccentricity at roughly fixed $q$, scattered bodies tend to fall back toward their pericentres rather than circularize at large $a$. In the sharp drop-off case, no particles attain $q>10$\,au ; in the sloped (canonical) disk, $< 1\%$ reaches $q\simeq12$\,au ; and in the flat disk, $< 1\%$ achieve  $q\simeq15$\,au. Even when $a>20$, the small $q$ means the orbits remain vulnerable to loss. This behavior is also consistent with the two-planet, gas-rich disk experiments of \citet{Ribeiro2024}, who examined scattering during Jupiter–Saturn growth in an \citet{Raymond2017a}-like disk context. Although their study emphasizes asteroid-belt implantation rather than retention in the ice-giant zone, their reported outer-disk outcomes similarly show that essentially none are circularized onto stable orbits beyond 12\,au over Myr timescales.

Moreover, suppose the outer disk contained partial gaps or ring-like density drops beyond Saturn, such as those that might arise at ice lines. In that case, these results indicate there would be little chance for substantial outward transfer of CM-rich material. In such lower-density regions, aerodynamic drag and scattering events would be too weak to bridge the $\sim8.6$\,au boundary effectively, as evidenced by the gas-poor, sharp drop-off profile, where 0\% of the initial CM population can reach stable $e<0.4$ orbits in the 15-20 au zone. Hence, the presence of any pronounced gas deficit outside Saturn’s orbit would further limit CM-like objects from migrating into the Uranus–Neptune zone, reinforcing the idea that distinct ring structures in the disk could isolate compositional reservoirs until another core-forming event reshaped the local dynamics.

Overall, even in the most gas-rich scenario, only a modest fraction of planetesimals survive outward scattering with sufficiently low eccentricity to remain long-term ($<$2\%). Most remain on highly eccentric paths or else are lost from the system, implying that a truly robust implantation of large CM-like bodies into the Uranus–Neptune region requires a combination of high disk density and additional dynamical events (e.g., interactions with later-forming embryos). Moreover, the very gas levels required to help particles remain would also enhance Type-I migration of embryos, further pulling pericentres inward and working against durable implantation. In sum, only exceedingly high outer-disk gas would have produced any outward retention at all, and even then at a percent-level at best; the contribution of CM-like material to the Uranus–Neptune formation zone and the present-day TNO population would therefore be negligible.

Taken together, these findings confirm that the two-planet configuration allows Saturn to scatter a significant fraction of CM-like bodies both inward to the asteroid belt and also outward into the 10–20\,au region, though whether or not their orbits will circularize depends on the amount of gas remaining in the Uranus-Neptune formation zone. Neither the precise growth rate of Saturn nor the final integration time alters the distribution of scattered planetesimals. Once the scattering window closes around $2.5 \times 10^5$ years, the fate of each particle is, for all practical purposes, sealed. We estimate planetesimal-planetesimal collisions in Annex~\ref{sec:coll}.

\subsection{Three planet scenario}

In the three-planet scenario, we extend the two-planet model by adding an ice-giant core at 9.6\,au, which grows from $1,M_\oplus$ to an approximately Uranus-mass ($\sim$15$M_\oplus$) over the prescribed $\tau_\text{growth}$. Whether this core represents a hypothetical `Ice~1' planet postulated by recent Nice Model variations \citep{Deienno2017} or an early Uranus-analog would require migration to match the current orbital configuration of the ice giants. We initialize planetesimals uniformly between 7\,au and 11\,au to capture the broader region where multiple planetary embryos could form.

The additional core’s dynamical impact is qualitatively similar to Saturn’s. It excites eccentricities for planetesimals in the vicinity of its orbit, scattering them along constant-Tisserand pathways. In the flat disk model (where Jupiter and Saturn carve wide gaps), the ice giant’s net inward drift remains moderate, and its scattering zone closely mirrors that of Saturn in the two-planet case. By contrast, since the sharp drop-off model sees intense inward migration due to the high amount of gas in the inner system, this core even manages to push Jupiter inward by nearly 1\,au, demonstrating how strongly the gas reservoir can affect overall giant-planet architecture.

We observe a somewhat stronger dependence on the growth timescale when the additional core is present, as demonstrated in Fig.~\ref{fig:3P_Imp}. Rapidly forming ice-giant cores reach higher masses earlier, thereby scattering more distant planetesimals and promoting greater orbital reshuffling at higher heliocentric distances. In slower-growth runs, the ice giant’s milder migration exerts comparatively less perturbation on bodies beyond $\sim10.4$\,au. Nevertheless, once scattered planetesimals do enter the ice giant’s sphere of influence, they show a similar pattern of partial circularization to that seen in the two-planet case (Fig.~\ref{fig:3P_AE}). Notably, objects initially beyond Saturn’s scattering limit ($\sim8.6$\,au) can now be pushed deeper into the Uranus–Neptune zone and beyond ($>30$\,au) with eccentricities $e<0.4$ in the $15-20$\,au range. However, their $q$s remain low, none exceeding $q>20$\,au for any gas profile.

Overall, the presence of this third core modestly widens the region of active scattering compared to the two-planet scenario, particularly under high-gas conditions and rapid core growth. At the same time, once again, objects forming even further out typically remain out of reach unless additional planetary embryos emerge or the local gas density remains sufficiently high for longer, thus allowing repeated scattering and aerodynamic damping to take hold.

\subsection{Reservoir contamination}

Even if only a small fraction of CM-like planetesimals can be scattered outward and circularized, it is useful to compare that contribution to the scale of the reservoir it would enter. In our working picture, the CM source region is comparatively narrow (7-11\,au), whereas the CI-like source region is typically placed farther out and spans a broader annulus (15-25\,au; \citealt{Hopp2022}), possibly already containing a native mass of order $M_{\rm CI}\sim20\,M_\oplus$. Adopting the CM mass budget inferred in \citet{Anderson2025} for loose planetesimal material between Saturn and Uranus ($M_{\rm CM,tot}\sim1\,M_\oplus$), our measured upper limit of ${\lesssim}2\%$ (two-planet) to ${\lesssim}4\%$ (three-planet) of 100-km bodies attaining long-lived, low-$e$ orbits beyond Saturn implies at most $M_{\rm CM}\lesssim 0.04\,M_\oplus$ reaching the 15-25\,au region, i.e. a mass fraction 
${\lesssim}2\times10^{-3}$ once diluted into the native ring. 

Such a contribution is too small to measurably affect the observed CM-CI dichotomy in the asteroid belt; any CM material that briefly resided in the outer ring would be both rare and, if it experienced outer-disk processing, potentially difficult to recognize unambiguously in the present-day, implanted population.

\subsection{Comparison with the five planet scenario}

\begin{table}
\caption{Distribution of planetesimals initially formed between $7<a_\text{init}<11$\, au at $t = 300\,\mathrm{kyr}$ for the five planet scenario from \citet{Anderson2025}, as percentages relative to the initial population of test particles.}
\label{tab:5Pl}
\centering
\begin{tabular}{lcccc}
\hline\hline
Planet  & $\tau_\mathrm{G}$ & $10$$<$$a$$<$$15$\,au & $15$$<$$a$$<20$\,au & $a$$>$$20$\,au \\ 
conf. &  & $e<0.4$ & $e<0.4$ &  $e<1$ \\
\hline
\multicolumn{5}{l}{\vrule height 10pt width 0pt Flat Gas Profile} \\
Wide    & $5\times10^5$ & 38.20 & 11.24 & 1.69 \\
Tight   & $5\times10^5$ & 39.45 &  9.54 & 1.73 \\
\hline
\multicolumn{5}{l}{\vrule height 10pt width 0pt Sloped Gas Profile} \\
Wide    & $1\times10^5$ & 11.18 &  9.68 & 4.87 \\
        & $5\times10^5$ & 26.63 &  8.93 & 3.68 \\
        & $1\times10^6$ & 36.42 &  6.03 & 2.87 \\
Tight   & $1\times10^5$ & 10.90 & 11.13 & 4.51 \\
        & $5\times10^5$ & 42.24 &  4.13 & 2.88 \\
        & $1\times10^6$ & 33.83 &  5.62 & 3.10 \\
No Mig. & $1\times10^5$ & 50.81 &  0.95 & 1.80 \\
\hline
\multicolumn{5}{l}{\vrule height 10pt width 0pt Sharp Drop-Off Gas Profile} \\
Wide    & $5\times10^5$ & 49.30 &  0.12 & 0.12 \\
Tight   & $5\times10^5$ & 39.19 &  2.51 & 1.93 \\
\hline

\hline
\end{tabular}
\end{table}

In our previous \citet{Anderson2025} study, we explored a model in which five planetary cores formed simultaneously. Additional cores were placed for Uranus at 15.4\,au and Neptune at 20.3\,au in the `Wide' configuration (3:2, 3:2, 2:1, 3:2 resonance chain), and at 12.6\,au and 16.2\,au, respectively, in the `Tight' variant (3:2, 3:2, 3:2, 3:2 resonance chain) \citet{Deienno2017}. Apart from these added cores, the underlying gas profiles (flat, sloped, or sharp) and $\tau_\text{growth}$ were analogous to those presented in this work. A natural consequence of having gas present in the disk is that all planetary embryos migrated inwards and drastically reshaped the distribution of planetesimals. Whereas the previous paper focused on the delivery of planetesimals to \emph{within} 5\,au, we can also re-examine these scenarios from the perspective of potential CM-like bodies that survived beyond 10\,au. For consistency with our current work, we will only focus on planetesimals with the initial starting semi-major axis $a_\text{init}$ between 7 and 11\,au. These results are summarized in Table~\ref{tab:5Pl}. Notably, none of these runs produce long-lived, low-eccentricity objects on stable orbits beyond 20\,au.

\begin{figure*}
    \includegraphics[width=\linewidth]{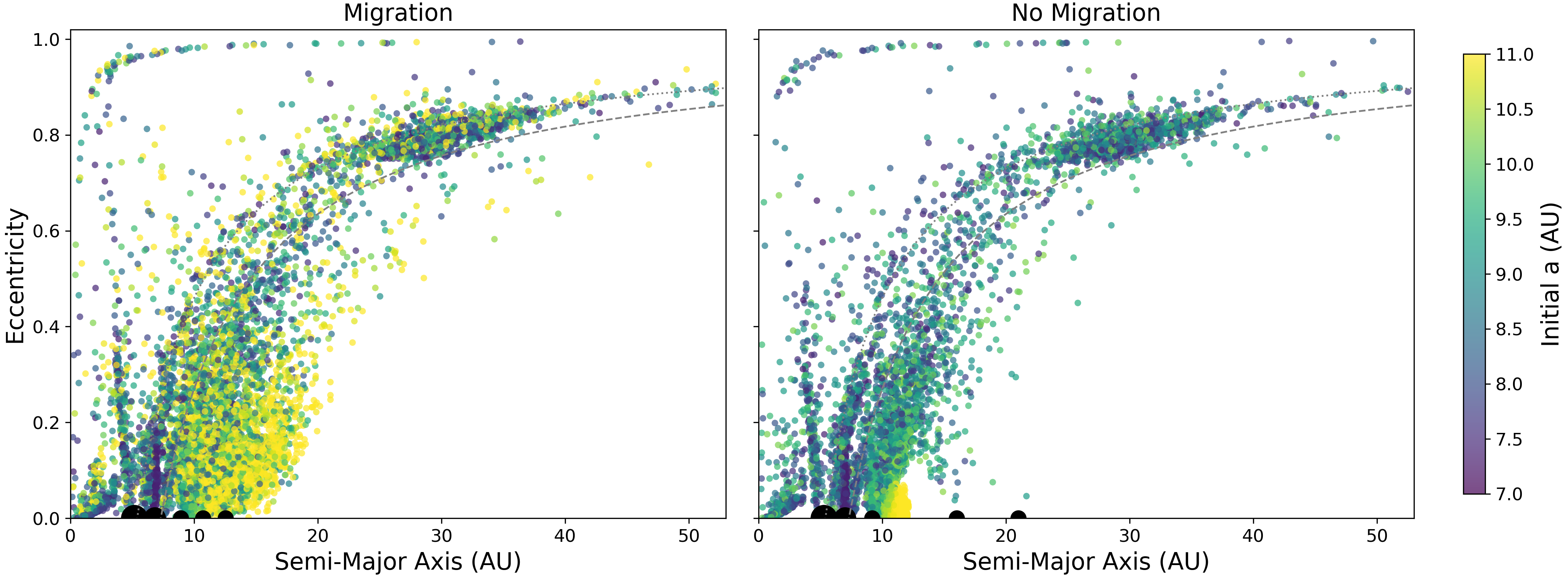}
     \caption{Same as Fig.~\ref{fig:2P_AE}, but for the five-planet models \citep{Anderson2025}. In both simulations, Uranus and Neptune are initialized in the `Wide' configuration. On the left, they are allowed to migrate naturally as they grow due to Type-I torque. On the right, the semi-major axes are fixed. Planetesimals are color-coded based on their formation location. In the simulations in which the ice giant cores are allowed to migrate, surviving planetesimals are circularized quickly.}
     \label{fig:5P}
\end{figure*}

What stands out in the five-planet scenario is that if Uranus’s and Neptune’s cores are allowed to migrate (Fig.~\ref{fig:5P}) while Saturn is still forming, while many planetesimals are ejected, a non-negligible fraction are quickly circularized into stable orbits between 15–20\,au within 300\,kyr. This is a clear departure from the two- and three-planet cases. In a flat gas profile, more than 10\% of the original 7–11\,au population survives on relatively stable orbits in the 15-20 au region, with Uranus and Neptune helping to capture and damp their orbits. 

To isolate the impact of migration, we ran a control `Wide, no-migration' simulation, in which the ice-giant cores’ semi-major axes were held fixed while they accreted (akin to \citealt{Nesvorny2024}). In this case, less than ${<}1\%$ of the 7–11\,au cohort is circularized between 15–20\,au by 300\,kyr. The contrast arises because migrating cores sweep across the 10–20\,au band, providing repeated, gentle encounters that lengthen planetesimals’ residence time at large $a$, so gas drag can damp $e$. Since more gas leads to faster core migration and stronger drag, this also leads to a larger stabilized fraction (up to 10\%), whereas suppressed migration yields minimal implantation (${<}1\%$).

\section{Discussion}\label{sec:discuss}

\paragraph{Confronting simulations with observations.}

Cycle 1-4 JWST/NIRSpec observations of $\sim$100 Centaurs and TNOs, primarily from the DiSCo-TNOs program \citep[e.g.,][]{Pinilla-Alonso2025, Licandro2025}, but also from complementary GO and GTO programs \citep[e.g.,][]{Emery2024, Holler2025}, have revolutionized our understanding of these populations by revealing remarkable compositional diversity in volatiles and organics across the outer Solar System. Cluster analyses consistently identify three distinct compositional groups among TNOs, $\text{H}_2\text{O}$-type Bowl-TNOs, $\text{CO}_2$-type Double-dip-TNOs, and organics-type Cliff-TNOs \citep{Holler2025, Pinilla-Alonso2025}, that correlate strongly with molecular abundances. In particular, no phyllosilicates were identified at the surfaces of these bodies, whose refractory component appears totally anhydrous based on JWST/MIRI observations \citep{Vernazza2025}.

If CM bodies are present among TNOs, they must be hidden in their interior. However, this scenario is not supported by near- and mid-infrared observations of Centaurs and TNOs. Mid-infrared observations show the systematic presence of anhydrous dust at the object surfaces, which appears similar to CP IDPs \citep{Vernazza2025}. Yet, CP IDPs differ in texture from CM chondrites, being chondrule-poor. A surface consisting of the anhydrous precursors of CM chondrites would somewhat resemble spectrally CO and CV chondrites, which show a mixture of matrix and chondrules; that is, an olivine band would be clearly present in the near-infrared range. Such a band is, however, not seen in the aforementioned surveys. Note that the presence of CI-type material is not seen either among current surveys of Centaurs and TNOs. However, CI-like material is present at the surfaces of some irregular satellites of Jupiter, which are thought to be, like the Jupiter Trojans, former TNOs \citep{Sharkey2025}.

To conclude, there is little to no evidence of CM-type material in the Kuiper Belt based on current surveys, fully in agreement with these simulations strongly suggesting inefficient mixing.

\paragraph{Less common carbonaceous chondrites.} These simulations confirm that large ($\sim$100\,km) planetesimals formed beyond $\sim$8.6\,au experience minimal dynamical disturbance if Saturn is the only currently forming core. This potentially leaves ample time for undisturbed planetesimals to undergo compositional changes or to be later scattered by an additional embryo that emerges in the outer disk, such as our three-planet model, in which case the undisturbed population exists ${>}10.4$\,au. Such intermediary populations might correspond to the less common carbonaceous chondrite subgroups (CK, CR, CH, CB) or even ungrouped carbonaceous materials, bridging the gap between the CM reservoir and a more distant CI-like region. Although this remains tentative, given the limited number of known large asteroids associated with these types. Current observations do not provide strong support for or against this possibility.

\paragraph{Sequential formation of giant planets.} The five-planet case highlights why coeval formation is difficult to reconcile with the observational constraints. When multiple ice-giant cores grow in a gas-rich disk, their Type-I migration and repeated encounters can raise the fraction of CM-like bodies temporarily “parked” at 15–20\,au to $\sim$5–10\%. However, this parking is inseparable from the same gas-rich, migratory phase that also promotes contemporaneous delivery from trans-Saturnian reservoirs. In that limit, CM-like and CI-like material are redistributed into the asteroid belt under similar disk conditions and at similar epochs, which would tend to erase the distinct implantation signatures inferred from the present-day radial distributions of CM- and CI-like asteroids.

Avoiding this outcome would require that the CI source region lies well beyond the dynamically accessed zone during the gas-rich phase, implying either a much more distant (${\gg}20\,{\rm au}$) CI reservoir or substantial intermediate reservoirs between CM and CI, neither of which is strongly supported by current constraints.
These results thus favor a sequential giant-planet formation model: Saturn’s formation and scattering of CM-like bodies likely occurred before Uranus and Neptune reached significant mass, or began migrating. This sequence of events preserves the distinct implantation profiles of CM and CI groups in the asteroid belt. It is also consistent with models where giant planets form one generation at a time in pressure maxima \citep[e.g.,][]{Lau2024}, with Saturn’s gap first halting inward flux of solids and only later a new structure emerging beyond it to seed the ice giants.

\paragraph{Arguments against implantation of CI-like bodies.}

Any interpretation based on implantation efficiencies must also be consistent with simple mass budgets. If the CM source region contained of order $M_{\rm CM}\sim1\,M_\oplus$ of leftover planetesimal mass, then an efficient inward delivery of $\sim$5–10\% would correspond to $\sim$0.05–0.1\,$M_\oplus$ of CM-like material implanted in the asteroid belt. By contrast, adopting a trans-Saturnian CI source region with $M_{\rm CI}\sim20\,M_\oplus$, a $\sim$2–5\% delivery efficiency would imply $\sim$0.4–1\,$M_\oplus$ of CI-like material. This would yield a CI:CM implanted mass ratio of 10:1 rather than the 1.1:1 observed today.

This not only differs substantially from the near-unity ratio but also far exceeds the current total main-belt mass \citep[$\sim4\times10^{-4},M_\oplus$][]{Pitjeva2018}. Some works \citep[e.g.,][]{Nesvorny2024} thus assume additional mechanisms, preferentially destroying CI-like bodies, for instance, planetesimal ablation by gas.
However, this mitigation requires sufficient gas to operate, and again reintroduces the problem
with explaining differing distributions of CM- vs. CI-like bodies.

\paragraph{Small (1--10\,km) planetesimals.}\label{sec:smallersize}

Although these simulations focus on 100\,km-size planetesimals, the conclusion that CM-like material is difficult to implant in the ice-giant region is not sensitive to this choice. \citet{Raymond2017a} found that such smaller planetesimals can more readily circularize at moderate distances when scattered outward, but do not necessarily extend their orbits as far as larger bodies. For small ($\sim$1–10\,km) bodies, their shorter gas-drag timescales allow them to lose orbital energy before they are fully ejected; consequently, many of them remain in stable orbits over $\sim$10\,au, though none reach beyond $\sim$15\,au. Meanwhile, 10\,km objects behave somewhat similarly to 100\,km bodies in their ultimate radial extent—often maxing out near $\sim$20\,au, but experience more effective eccentricity damping than our reference case.

Such objects would be particularly susceptible to reprocessing, collisional grinding, or re-accretion onto later-forming planetary embryos. Future work incorporating explicit size distributions\textemdash including pebbles and dust\textemdash will be crucial for assessing whether such processes could help explain the compositional diversity of distant objects and the persistence of chondritic materials in orbits beyond Saturn.

\section{Conclusions}\label{sec:conclusions}

We show that large (100-km) CM-like planetesimals formed at the outer edge of Jupiter’s gap are efficiently scattered by Saturn but only weakly implanted beyond Saturn’s orbit. In two-planet runs, Saturn clears a narrow annulus around its path and sends a few percent (4-6\%) inward to the asteroid belt, while bodies that formed beyond $\sim$8.6\,au largely retain semi-major axes close to their birth locations, experiencing little scattering or long-range transport on the timescales explored, and could be considered dynamically sheltered. Adding a single ice-giant core at $\sim$9.6 au modestly extends Saturn’s reach, and while up to 10\% are scattered inwards, outward implantation of their bodies into long-lived orbits beyond 10\,au remains intrinsically inefficient: once scattered to high-e, gas drag damps toward pericentre rather than circularizing at larger radii. Even in gas-rich cases, only up to $<4$\% of 100 km objects achieve stable orbits in the Uranus-Neptune zone.

When multiple ice-giant embryos co-form and migrate inward, circularization in the 15–20 au region becomes faster and more common, up to 10\%. The embryos are subject to Type-I migration, which continues to reduce their orbits and to steer the planetesimals along to lower $q$, so even in this favorable case it remains difficult to `park' substantial CM-like mass on wide, long-lived orbits at large distances. Conversely, if extensive CM contamination of the outer disk had occurred at the same epoch as CI formation, subsequent outward migration during the instability would have imprinted similar distributions of CM and CI bodies in the asteroid belt, contrary to observations. A key observational constraint is the lack of CM-like spectra among trans-Neptunian objects in current surveys. Any CM-like population implanted into the 15-20\,au region would also represent only a small contaminant (${\lesssim}0.2\%$) of the much broader outer reservoir, often taken to span $\sim$15-25\,au, and only a further subset (${\lesssim}0.05\%$) would be expected to reach the main belt during later phases, with an upper limit of ${\lesssim}2\times10^{-3}M_\oplus$.

Taken together, these results highlight that attempts to place CM-like material far beyond Saturn’s orbit must contend with either substantial gas availability or additional planetary embryos that can shepherd planetesimals into stable orbits. In this picture, pressure bumps and gaps act as semi-permeable barriers that preserve compositional rings with limited inter-ring mixing. The absence of implantation of CM-like bodies even under gas-dense conditions, together with the distinct CM and CI distributions in the main belt and the lack of CM-like spectra among TNOs \citep{Vernazza2025}, strongly indicates that the CM and CI source reservoirs were spatially and temporally isolated. 

\begin{acknowledgements}
The work of S.E.A. has been supported by the Institut Origines of Aix Marseille University, and the work of P.V. by CNES, CNRS/INSU/PNP, and the Institut Origines. The work of M.B. has been supported by the Czech Science Foundation (grant number 25-16507S).
\end{acknowledgements}

\bibliographystyle{aa}
\bibliography{bibliography.bib}

\newpage

\begin{appendix}

\onecolumn

\section{Collisional Efficiencies}\label{sec:coll}

\subsection{Planetesimal-planetesimal collisions.} 

Our simulations neglect planetesimal-planetesimal collisions, which could, in principle, modify implantation outcomes by fragmenting bodies into smaller sizes that couple more strongly to gas drag. This effect is particularly relevant during phases of giant-planet growth and migration: \citet{Carter2020} demonstrated that giant planet migration can drive a substantial number of high-velocity, potentially disruptive collisions. In their study, collisional processing can significantly reshape the population ultimately delivered to the inner Solar System and asteroid belt, even when planets are not strongly evolving.

We estimate mutual collision rates among the outwardly scattered CM-like bodies using a particle-in-a-box approximation. We first identify the subset of survivors satisfying $a>15$\,au and $e<0.4$ at 300\,kyr for each simulation $s$, and convert these fractions $f_s$ into a physical number $N_s$ of bodies by adopting the total CM reservoir mass of $M_{\rm CM} \approx 1\,M_\oplus$:

\begin{equation}
    N_s = f_s\frac{M_{\rm CM}}{\frac{\pi}{6}\rho D^3}
\end{equation}

The corresponding number density $n_s$ is defined as $n_s = N_s/V$, where $V$ is the volume of the annulus sampled by these orbits, approximated as a 10\,au wide annular torus centered on $a=20\,{\rm au}$. The collisional cross section is defined as $\sigma = {\pi\over 4} D^2$. We take the characteristic relative velocity $v_{\rm rel,s} = v_K\sqrt{e_s^2+i_s^2}$ where $e_s$ and $i_s$ are the mean values of $e$ and $i$, respectively, of the surviving particles in each simulation. The resulting collisional timescale is:

\begin{equation}
    t_\mathrm{coll}=\frac{n_s}{\sigma v_\mathrm{rel,s}}
\end{equation}

For all disk profiles tested, we obtain $t_\mathrm{coll}>6\times10^{9}$\,yr, longer than the age of the solar system. This equates to a per-object collision probability of 
$P\simeq 1-\exp(-300\,\mathrm{kyr}/t_\mathrm{coll})<5\times 10^{-3}$\% over 300 kyr for each planetesimal, with the highest probability being in the flat-gas-profile case. Whether the simulations had two or three planets has less of an impact on the collisional probability than the gas profile itself, with sharp-drop off profiles having $0-10^ {-5}$\% probability, sloped profiles having $10^{-4}$\% probability, and flat profiles having $10^{-3}$\%. In the specific context of the outer disk, collisions are unlikely to alter our main conclusions: Across all of our runs, no bodies are circularized onto stable orbits at $a>20$\,au with $e<0.4$, and only up to $<4\%$ reach the $15$-$25$\,au region on low-eccentricity orbits in our most generous case. In other words, while collisions may strongly affect the size distribution and implantation efficiency inside $\sim$10\,au (as emphasized by \citealt{Carter2020}), the scarcity of long-lived CM-like material at large heliocentric distances in our simulations makes collisional reprocessing an inefficient pathway for enhancing CM-CI mixing in the ice-giant zone.

\subsection{Collisions with the native disk.}

Collisions with a pre-existing `native' CI outer-disk population could be more frequent if such a population had already formed. To place an upper bound on this channel, we assumed a planetesimal disk of total mass $M_\mathrm{CI}=20M_\oplus$ between 15 and 25\,au \citep[e.g.,][]{Hopp2022, Nesvorny2024}. Using the same approach as above, we change the approximation of $v_{\rm rel} = v_K\sqrt{e_s^2+e_{\rm CI}^2+i_s^2+i_{\rm CI}^2}$. We take an eccentricity of $e_{\rm CI}=0.05$ and inclination of $i_{\rm CI}=3^\circ$ for the native disk.  

In this case, the resulting collisional timescales all lie within a narrow range:  we estimate $t_\mathrm{coll,CM-CI}\sim (1.3-1.8) \times 10^7$\,yr, so over 300\,kyr, the per-object collision probability is $P\simeq 1-\exp(-300\, \mathrm{kyr}/t_\mathrm{coll})\approx 2\%$. Such impacts would likely be highly disruptive at the relative velocities inferred here ($v_{\rm rel} =2-3~{\rm km.s}^{-1}$), and therefore would tend to destroy rather than reprocess the material. Taken together, both the small implanted fraction and the long CM–CM collision timescales indicate that outward contamination by large CM-like bodies is dynamically and collisionally inefficient, and would be unlikely to alter the bulk character of a CI-dominated outer reservoir measurably.

We note that \citet{Carter2020} also found that outcomes can differ in configurations where Jupiter and Saturn form substantially farther out (e.g., beyond $\sim$10\,au), even when adopting disk conditions comparable to those explored here. Such architectures increase the overlap between the scattering region and the outer disk, potentially raising collision rates. Since our nominal setup places Jupiter and Saturn in a more compact configuration, our results imply that collisional grinding is not a limiting factor for the outer-disk mixing problem: the bottleneck is the dynamical inefficiency of placing large CM-like bodies onto low-eccentricity, long-lived orbits beyond Saturn in the first place.

\newpage
\section{Supplementary figures}

\begin{figure*}[ht!]
\centering
    \includegraphics[width=0.9\linewidth]{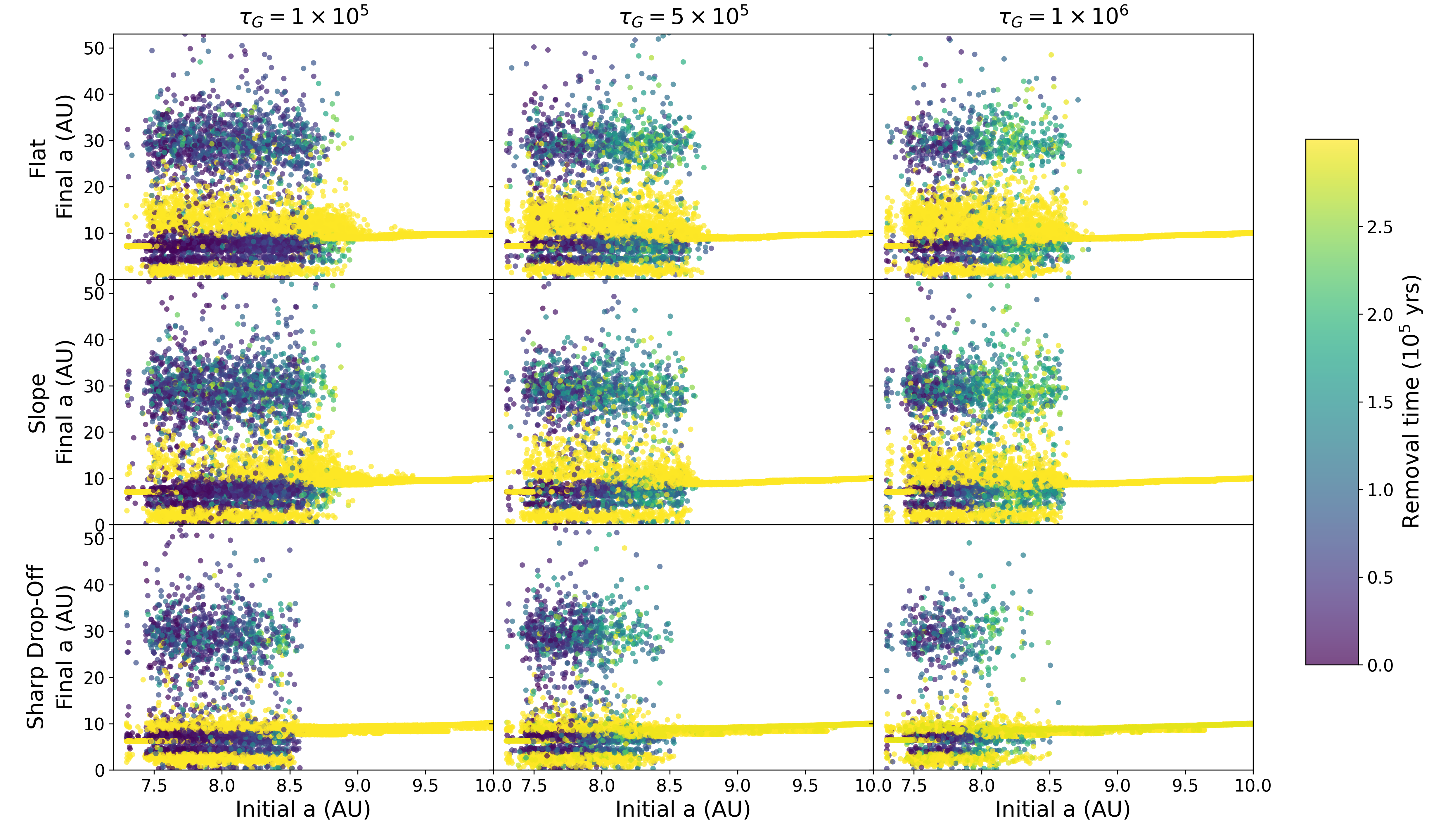}
    \label{fig:2P_Init}
    \includegraphics[width=0.9\linewidth]{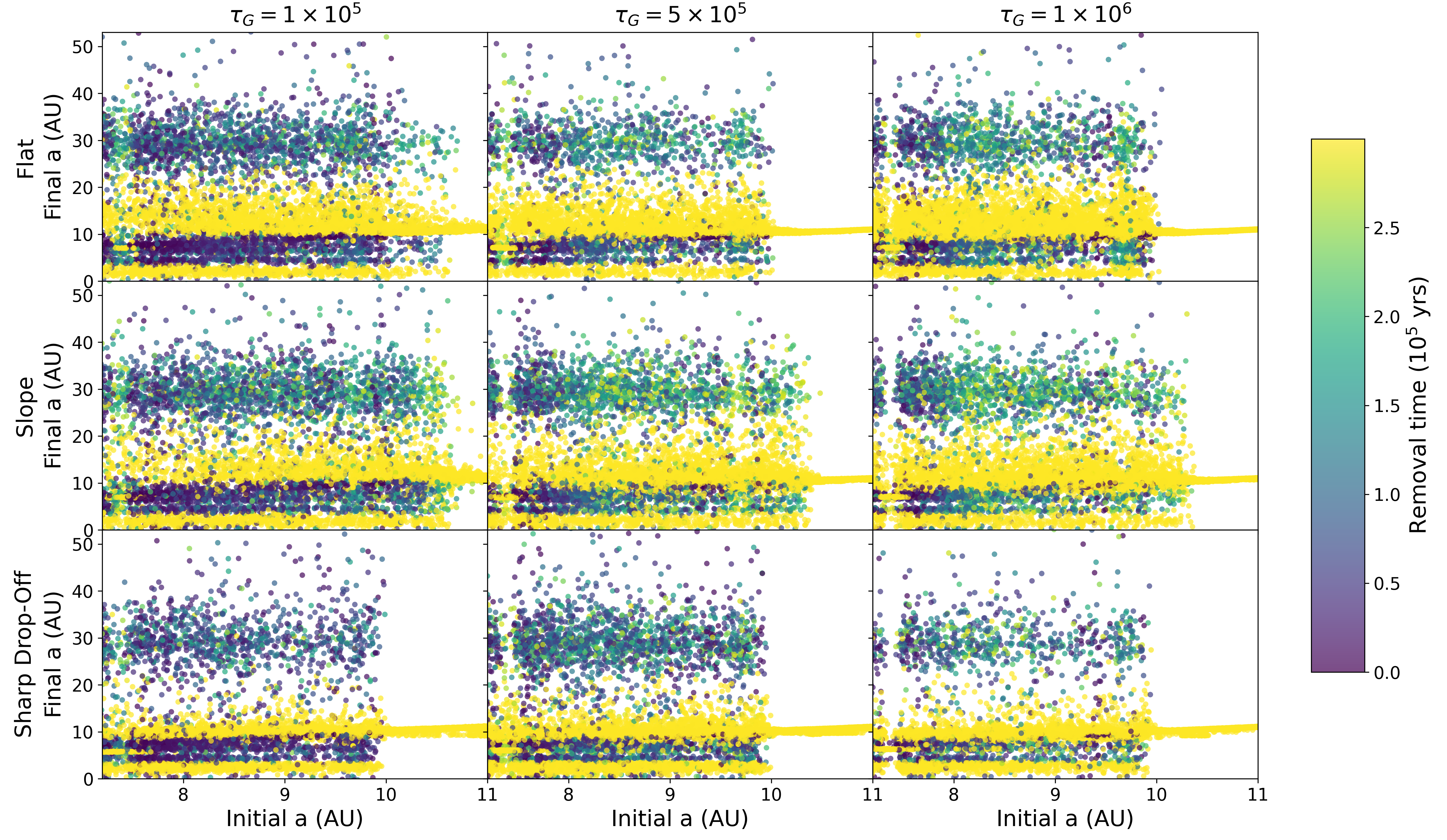}
    \caption{Semi-major axis of planetesimals at $t = 300\,\mathrm{kyr}$ as a function of their initial formation zone for each gas profile and Saturn growth timescale for our two-planet model (top) and three-planet model (bottom). The color represents the moment the test particle was lost from the simulation, with cooler colors representing earlier loss and yellow for those that remain. The extent of Saturn's `reach' is nearly identical across all simulations. The objects outside this reach, the somewhat horizontal yellow bar, will retain their initial semi-major axis, while planetesimals formed around Saturn may `leapfrog' over them to higher heliocentric distances.}
    \label{fig:3P_Init}
\end{figure*}

\end{appendix}

\end{document}